\documentclass[aps,amsmath,prc]{revtex4}
\usepackage{epsfig}
\usepackage{color}
\usepackage{hyperref}
\newcommand{\dq}{\mbox{d\hspace{-.55em}$^-$}}
\newcommand{\mn}{m_\text{\tiny N}}
\newcommand{\ecm}{E_\text{\tiny c.m.}}
\newcommand{\app}{a^C(pp)}
\newcommand{\ann}{a_s(nn)}
\newcommand{\anps}{a_s(np)}
\newcommand{\anpt}{a_t(np)}
\newcommand{\nt}{$t$-$n$}
\newcommand{\ant}{a_{s/t}(t\text{-}n)}
\newcommand{\ants}{a_s(t\text{-}n)}
\newcommand{\antt}{a_t(t\text{-}n)}
\newcommand{\phe}{$^3$He-$p$}
\newcommand{\aphe}{a^C_{s/t}(^3\text{He-}p)}
\newcommand{\aphes}{a^C_s(^3\text{He-}p)}
\newcommand{\aphet}{a^C_t(^3\text{He-}p)}
\newcommand{\ctni}{C_{3\text{NI}}}
\newcommand{\eftnopi}{\mbox{EFT$(\not \! \pi)$}}
\begin{document}
\title{Zero-energy neutron-triton and proton-Helium-3 scattering\\with \eftnopi}

\author{Johannes Kirscher}
\affiliation{Department of Physics and Astronomy, Ohio University, Athens, Ohio 45701, USA}

\begin{abstract} \vspace*{18pt}
Model-independent constraints for the neutron-triton and proton-Helium-3 scattering lengths are calculated with a leading-order interaction
derived from an effective field theory without explicit pions. Using the singlet neutron-proton scattering length, the deuteron, and the triton binding
energy as input, the predictions $\ants=9.2\pm2.6~$fm, $\antt=7.6\pm1.6~$fm, $\aphes=3.6\pm0.32~$fm, and $\aphet=3.1\pm 0.23~$fm are obtained.
\par
The calculations employ the resonating group method and include the Coulomb interaction when appropriate.
The theoretical uncertainty is assessed via a variation of the regulator parameter of the short-distance interaction from $400~$MeV to $1.6~$GeV.
The phase-shift and scattering-length results for the proton-Helium-3 system are consistent with a recent phase shift analysis and with model calculations.
For neutron-triton, the results for the scattering lengths in both singlet and triplet channels are significantly smaller than suggested
by R-matrix and partial-wave-analysis extractions from data.
\par
For a better understanding of this discrepancy, the sensitivity of the low-energy four-body scattering system to variations in the neutron-neutron and proton-proton
two-nucleon scattering lengths is calculated. Induced by strong charge-symmetry-breaking contact interactions, this dependence is found insignificant.
In contrast, a strong correlation between the neutron-triton scattering length and the triton binding energy analogous to the Phillips line is found.
\par
\end{abstract}
\maketitle
%----------------------------------------------------------------------------------------------------------------------------------------------------
\section{Introduction}
The refinement of numerical methods (\textit{e.g.}~\cite{gfmc,lit,hmh-rrgm,fy}) and technological progress made the application of high-precission
models of the nucleon-nucleon and three-nucleon interaction (\textit{e.g.}~\cite{uix,bonn}) to the four-body system feasible more than a decade ago.
By now, the numerical approximations to the solution of the four-body problem are accurate enough to relate the differences
in the predicted observables solely to the differences in the nuclear force models.
Parallel to this development, the effective field theory (EFT) formalism has been successfully applied to an abundance of nuclear processes,
mostly in smaller systems like the deuteron and triton (see \textit{e.g.}~\cite{eft-mam-rev,rev-pionless-platter} for reviews).
This formalism has also been applied in~\cite{platter-4bdy-lo,bauchpinsel} to understand the Helium-4 nucleus as a universal consequence of
peculiar features of the two- and three-nucleon systems namely unnaturally large scattering lengths and associated shallow bound states.
The results of this work establish for the first time an analogous correlation between the four-nucleon scattering system and its subsystems with a nuclear EFT.
Furthermore, the absence of bound four-body states in the investigated systems distinguishes this analysis from previous EFT approaches to the four-body system.
\par
Two characteristics of the EFT approach make its application here especially advantageous:
first, its ability to predict four-nucleon scattering observables with theoretical error bars, and second,
the extent to which variations in two- and three-body observables flow through into four-body observables can be calculated.
Knowledge of both the theoretical uncertainty and the dependence on observables in smaller subsystems is required, first, for testing the effective range parameters of the four-nucleon
system extracted in a phase shift analysis (PSA) on their consistency with two- and three-nucleon data.
Second, it explains the discrepancy in predictions for those parameters obtained from nuclear force models.
\par
The experimental and theoretical status of low-energy proton-Helium-3 (\phe) scattering is stated briefly to specify the focus of this work
which is on the singlet/triplet S-wave scattering lengths $\aphe$~as a diagnostic for the nuclear interaction in this system.
The latest experimental extraction of this value via PSA~\cite{phe-psa} yielded: $\aphes=11.1\pm0.5~$fm and $\aphet=9.05\pm0.15~$fm. Both values differ
significantly from previous PSAs~\cite{phe-knutson,phe-data8} (see table~\ref{tab.num-unc}), reflecting the relatively small database especially
regarding cross-sectional data below $1~$MeV and the associated uncertainty in the extrapolation to zero energy.
Therefore, na\"ively one would expect that new data could again shift the experimental scattering lengths by more than the quoted statistical errors.
\par
Theoretical predictions are at present inconclusive about the precise values of the four-body scattering lengths, too.
On the one hand, the model predictions for $\aphe$~show a strong sensitivity
to the Helium-3 binding energy $B(^3\text{He})$ obtained with the respective model, and on the other hand,
a less significant sensitivity to strong isospin-symmetry-violating interactions and thereby to the proton-proton scattering length $\app$.
This is manifest in the $\aphe$~values of $v_{14}$ and $v_{18}+$UIX, where the models differ in $B(^3\text{He})$,
and in those of $v_{14}$ and $v_{18}$, where the former respects and the latter violates charge symmetry.
In table~\ref{tab.num-unc}, a selection of recent experimental extractions and theoretical predictions is shown.
\par
To understand this discrepancy amongst theoretical and experimental values the effective field theory approach to the nuclear interaction is chosen here.
The aforementioned feature of an EFT calculation to provide theoretical error bars constrains the observables $\aphe$~to certain intervals.
Specifically, the appropriate EFT for very low-energy phenomena in nuclear systems, formulated without explicit pions (\eftnopi), is employed.
Its usefulness in describing even three- and four-nucleon systems is known by now (see \textit{e.g.}~\cite{rev-pionless-platter} for a review).
With this interaction theory, the resonating group method (RGM~\cite{hmh-rrgm}) is used to solve the few-body problem numerically, in which course it is shown
below that a leading-order (LO) calculation in \eftnopi~is consistent with the new PSA extraction for $\aphe$~and with high-precision-model predictions for these quantities.
\par
In addition, the EFT approach is used for a variation of the two-body scattering lengths without losing consistency with other low-energy data.
A charge-dependent interaction is introduced to shift $\app~$ and the neutron-neutron ($\ann$) scattering length relative to their values obtained in the SU(4) symmetric theory.
In that case, $\ann~$ is equal to the singlet neutron-proton scattering length ($\anps$).
It is thus possible to assess the sensitivity of the four-nucleon scattering lengths to $\ann$~independently
from a change in the triton binding energy $B(t)$ accompanying the variation of $\ann$.
Regarding experimental uncertainties, the \nt~system is more interesting than \phe, because the experimental knowledge about $\ann$~is comparable to $\aphe$,
\textit{i.e.}, conflicting extractions stem from insufficient data.
The proton-proton scattering length, in comparison, is quite accurately determined.
It is therefore of interest to quantify the effect of a different value for $\ann$~on the prediction for $\ant$ while employing
a proper renormalization of the three-nucleon interaction (TNI) so that $B(t)$ stays at the physical value.
In this way, the change in $\ant$~in all experimentally admissible scenarios for $\ann$~is explored without losing consistency with other data.
\par
The results show two similarities to the neutron-deuteron ($n$-$d$) system as investigated in~\cite{n-d-wit-nog}:
an underprediction of $\ant$~compared to its experimental value, and a correlation of $\ants$~to the binding energy of the triton (for $n$-$d$ called Phillips line~\cite{phillips-line}).
This work then establishes the four-nucleon scattering system as a universal consequence of the two- and three-body system,
when so far, only a bound four-nucleon system in the Helium-4 channel was identified as an emergent property of the shallow two- and three-body states (see \textit{e.g.}~\cite{platter-tjon}).
\par
This report continues with a summary of the employed nuclear interaction based on \eftnopi~and the specification of the numerical parameters used in the RGM calculation.
The following presentation of the results for $\aphe,~\ant$, and the correlation between $\ants$ the triton binding energy - the
analog of the Tjon line~\cite{tjon} for a scattering observable - includes a discussion about the reliability of the error estimates, which is crucial for the conclusion.
%----------------------------------------------------------------------------------------------------------------------------------------------------
\section{Formulation}
%--
The EFT without explicit pions at LO yields a potential
\begin{eqnarray}\label{eq.pot-vertex}
\hat{V}&=&\sum_{i<j}^Af_\Lambda(\vec{r}_{ij})\left(C_S+C_T\vec{\sigma}_i\cdot\vec{\sigma}_j\right)+\Big[\left(\frac{e^2}{4\vert\vec{r}_i-\vec{r}_j\vert}+
C^{\text{\tiny pp}}f_\Lambda(\vec{r}_{ij})\right)\left(1+\tau_{i,3}\right)\left(1+\tau_{j,3}\right)\nonumber\\
&&+C^{nn}\left(1-\tau_{i,3}\right)\left(1-\tau_{j,3}\right)f_\Lambda(\vec{r}_{ij})\Big]\frac{1}{2}\left(1-\vec{\sigma}_i\cdot\vec{\sigma}_j\right)+
\sum\limits_{\stackrel{i<j<k}{\text{cyclic}}}^Af_\Lambda(\vec{r}_{ij})\cdot f_\Lambda(\vec{r}_{jk})\ctni\,\vec{\tau}_i\cdot\vec{\tau}_j\;\;\;,\nonumber\\
\end{eqnarray}
where a vertex with low-energy constant (LEC) $\tilde{C}$ was regulated via $\tilde{C}\to e^{-(\vec{p}-\vec{p}')^2/\Lambda^2}C$ with $p(p')$ denoting the center-of-mass momentum
of the incoming (outgoing) nucleon pair $ij$ yielding the radial dependence $f_\Lambda(\vec{r}_{ij}):=\left(\frac{\Lambda^3}{8\pi^{3/2}}\right)e^{-\frac{\Lambda^2}{4}\vec{r}_{ij}^2}$ on the
relative coordinate $\vec{r}_{ij}$.
The (iso)spin matrices ($\vec{\tau}$)$\vec{\sigma}$ with indices specifying the nucleon $i$ and Cartesian-z-component $3$
project onto spin singlet and triplet. The Coulomb interaction with electric charge $e$ represents the long-range part of the interaction.
For a charge-independent interaction, the three LECs, $\lbrace C_S,C_T,\ctni\rbrace$, were fitted to $\anps$, the deuteron binding energy
$B(d)$, and $B(t)$. When charge-independence/symmetry-breaking interactions are included, all five LECs $\lbrace C_S,C_T,C^{nn},C^{pp},\ctni\rbrace$ were fitted to
$\lbrace\anps,\app,\ann,B(d),B(t)\rbrace$. The strong isospin-violating terms are suppressed relative to the isoscalar LO interaction and are
presumably of the same order as the momentum dependent NLO isoscalar vertices~\cite{walz}. For the prediction of $\ant$ and $\aphe$, the inclusion of
$C^{nn,pp}$ is another probe of sensitivity to higher-order contributions in addition to the cutoff variation and analogous
to a replacement of $\anpt$ by $B(d)$ to fit $C_T$.
These two methods are used to map out the LO uncertainty due to suppressed short-range interactions: a variation of input data and a change of the regulator parameter $\Lambda$.
For $\Lambda$, values of $400,~800,~$and
$1600~$MeV were used. The lower bound was chosen in the vicinity of the natural breakdown scale of \eftnopi~while
the $\Lambda$ dependence up to $1.6~$GeV was found sufficient for an extrapolation to $\Lambda=\infty$.
\par
The particular incarnation of the RGM used here was shown to be accurate in its prediction of the four-body scattering system with
modern nuclear forces~\cite{hmh-4he}.
For details about the RGM the reader is therefore referred to~\cite{hmh-rrgm} where the method is introduced. Here, only the relevant parameters of the method are listed.
The model space used to obtain the results presented here includes two two-fragment channels for each four-body calculation, \phe, singlet-deuteron(\dq)-$nn$,
and \nt, \dq-$pp$ respectively.
In the triplet channels, only the \phe~and \nt~groupings contribute.
For LO accuracy in the interaction, it is sufficient to expand the $J^\pi=0^+$ channel
in pure S-waves, \textit{i.e.}, the orbital angular momenta on the two fragment-internal Jacobi coordinates of the
triton and Helium-3 and on the relative coordinate to the neutron or proton are zero.
\par
The numerical uncertainty due to the finite variational basis was assessed by using two different three-body bases ($80$ and $400$
dimensional); by a scaling of the $20$ Gaussian widths used to expand the relative wave function by a factor of $10$;
and by including distortion channels for a better approximation of the scattering state in the interaction region. The
four-body model space was deemed sufficient if the lowest eigenvalue within that space differed by less than
$60~$keV from the respective three-body binding energy. This guarantees that the basis satisfies the minimal condition of being large enough to form the bound three-nucleon state
with the fourth nucleon being free. Hence, it is expected that the basis is appropriate to expand asymptotic states.
For the accurate approximation of intermediate states, it was tested if further addition of distortion channels did affect the prediction for the scattering length by more than $0.01~$fm.
Once this level was reached, the basis was assumed to be complete for practical purposes.
\par
It is worth noting, that the underlying na\"ive expectation that even at LO neither \phe~nor \nt~sustain a bound four-body state is justified by the relatively large energy gap of about $5~$MeV from the \nt/\phe~threshold to the lowest resonant states. In contrast, a similar argument would fail in the Helium-5 system where it is not obvious why the observed shallow P-wave resonances should not become bound at LO in \eftnopi~given the accuracy of a LO calculation.
\par
The variational RGM solution for the S-matrix is obtained for energies below $\ecm=5~$keV and parameterized by a phase
shift $\delta$ via $S=e^{2i\delta(\ecm)}$. To extract a scattering length $a_s$, the standard effective range formula is
appropriate for \nt,
\begin{equation}\label{eq.ere-regular}
a_s=-\frac{1}{k\cot\delta(\ecm)}\;\;\;,
\end{equation}
while for fragments carrying charges $q_{1,2}$ its generalization (\textit{e.g.}~\cite{gen-ere})
describes the relative phase shifts in the presense of a long-range Coulomb potential
\begin{equation}\label{eq.ere-gen}
a^C=-\left(C^2_0(\eta)k\left[\cot\delta-i\right]+2k\eta H(\eta)\right)^{-1}\;\;\;\text{with}
\end{equation}
%--
\begin{equation}
C^2_0(x)=\frac{2\pi x}{e^{2\pi x}-1}\;\;;\;\;\eta=\frac{q_1q_2\alpha 3/4\mn}{2k}\;\;;\;\;k=\sqrt{3/4\mn\ecm}\;\;;\;\;H(\eta)=\frac{\Gamma'[i\eta]}{\Gamma[i\eta]}+\frac{1}{2i\eta}-\ln[i\eta]\;\;.
\end{equation}
%--
Formulas~(\ref{eq.ere-regular},\ref{eq.ere-gen}) were fitted to the respective RGM \eftnopi~predictions for the phase shifts where the optimal effective-range parameter $\aphe$/$\ant$ was found stable against inclusions of higher-order
terms in the expansions. Explicitly, even if $-\frac{1}{a}+\frac{r}{2}k^2$ was fitted to $k\cot\delta$ the result for $a$ did not shift by more than $0.05~$fm.
%----------------------------------------------------------------------------------------------------------------------------------------------------
\section{Results}
%--
First, the effective range parameters calculated with the RGM employing \eftnopi~interactions are shown in table~\ref{tab.num-unc}.
The predictions are compared to previous calculations selected to include models which yield the physical three-body binding energy
($v_{18}+$UIX, MT-I/III, and $v_{14}+$UIX), which only reproduce the two-nucleon sector accurately ($v_{14,18}$), and which break ($v_{18}$) or
conserve ($v_{14}$, MT-I/III) isospin symmetry.
The given theoretical uncertainties for the \eftnopi~predictions of this work are derived from cutoff- and two-body input-data variations.
The numerical RGM error was assessed independently and contributes a relatively small amount of $0.05~$fm to the total uncertainty.
For the central value, the EFT prediction corresponding to the smallest cutoff was chosen because then the interaction,
which was renormalized to $B(t)$, yields a Helium-3 binding energy very close to the datum. Increasing $\Lambda$ and
thereby admitting higher-momentum modes in the bound states leads to a stronger contribution from the repulsive Coulomb force
effectively lowering $B(^3\text{He})$.
\par
The error analysis discriminates between sensitivity of $\aphe$/$\ant$~to $a_{pp/nn}$ and $B(^3\text{He}/t)$ respectively.
A shift in the three-body binding energies, induced by
a change in the two-nucleon scattering lengths, is absorbed in the TNI as the triton is fixed to have its physical binding energy.
The effect on $\aphe$/$\ant$~due to a shift in $a_{pp/nn}$ is thereby found insignificant compared to the effect of a change in the binding
energy of the three-nucleon target. Specifically, the predicted value for $\ant$~changed by less than 1\%
when $C^{nn}$ is fitted to either of the two proposed experimental values for the neutron-neutron scattering length,
\textit{i.e.}, $\ann=-18.7\pm0.7~$fm~\cite{GO99,GO06} and $\ann=-16.1\pm0.4~$fm~\cite{HU001,HU002}, as long as the TNI
is used to fix $B(t)=8.48~$MeV. The model predictions are in agreement with this finding of a stronger dependence of four-nucleon scattering parameters on the
target's binding energy than on two-body scattering lengths.
%--
\begin{table*}
\renewcommand{\arraystretch}{1.1}
  \caption{\label{tab.num-unc}{\small Theoretical and experimental values for $\ant$~and $\aphe$~scattering lengths.
Predictions of charge-(as)symmetric interactions with and without three-body models are compared to phase-shift-analysis extractions based on different sets of data.}}
\scriptsize
\begin{tabular}{cccccc}
\hline\hline
\multicolumn{6}{c}{Theory}\\
$\aphes~$[fm] & $\aphet~$[fm] & $\ants~$[fm]& $\antt~$[fm] & interaction (method) & Ref.\\
\hline
 $11.5$ & $9.2$ & $4.10$ & $3.63$ & MT-I/III (FY) & \cite{laz-atn}\\
 $12.7$ & - & $4.28$ & $3.81$ & $v_{14}$ (FY) & \cite{laz-atn}\\
 $11.3$ & - & $4.04$ & $3.60$ & $v_{18}+$UIX (FY) & \cite{laz-atn}\\
 $11.5$ & $9.13$ & $4.06$ & $3.59$ & $v_{18}+$UIX (CHH) & \cite{nt-phe-kiev}\\
 $12.9$ & $10.0$ & $4.28$ & $3.73$ &$v_{18}$ (CHH) & \cite{nt-phe-kiev}\\
 -&-& $4.21$ & $3.54$ & $v_{14}+$UIX (CHH) & \cite{nt-phe-kiev}\\
 -&-& $4.32$ & $3.79$ &$v_{14}$ (CHH) & \cite{nt-phe-kiev}\\
 $9.2\pm2.6$ & $7.6\pm1.6$ & $3.6\pm0.32$ & $3.1\pm 0.23$ & \eftnopi~, $\Lambda\to\infty$ (RGM) & This work\\
\hline
\multicolumn{6}{c}{Experiment}\\
 $\aphes~$[fm] & $\aphet~$[fm] & $\ants~$[fm]& $\antt~$[fm] & dataset & Ref.\\
\hline
 $10.8\pm2.6$ & $8.1\pm0.5$   &-&-& \cite{phe-data8} & \cite{phe-data8,nt-phe-kiev,nt-carbonell}\\
 $15.1\pm0.8$ & $7.9\pm0.2$   &-&-& \cite{phe-data8,phe-data10,phe-data6}, solution 1 & \cite{phe-knutson}\\
 $7.2\pm0.8$  & $10.4\pm0.4$  &-&-& \cite{phe-data8,phe-data10,phe-data6}, solution 2 & \cite{phe-knutson}\\
 $11.1\pm0.5$ & $9.05\pm0.15$ &-&-& \cite{phe-data8,phe-data10,phe-data6,phe-data-fisher,phe-psa} & \cite{phe-psa}\\
-&-& $4.453\pm0.1$ & $3.325\pm0.016$ & \cite{rmat-tn} &\\
-&-& $4.98\pm 0.29$  & $3.13\pm 0.11$ & \cite{nt-exp} &\\
\hline\hline
    \end{tabular}
\end{table*}
%--
The EFT values for $\aphes~$ and $\aphet$~are consistent within error bars with model calculations which yield the physical three-nucleon binding
energies.
While the \eftnopi~results for \phe~agree also with the latest experimental numbers,
in the \nt~system, the \eftnopi~predictions are significantly smaller than suggested by models and R-matrix data.
The discrepancy to the models is explained below with the correlation between $B(t)$ and $\ant$~and thereby with significantly different three-body binding energies.
It is also interesting to note that for the difference $\ants-\antt$, all theoretical calculations, including this work, predict approximately the same number and hence collectively differ from the experimental extractions.
%--
\par
\begin{figure}
  \includegraphics[width=.7\columnwidth]{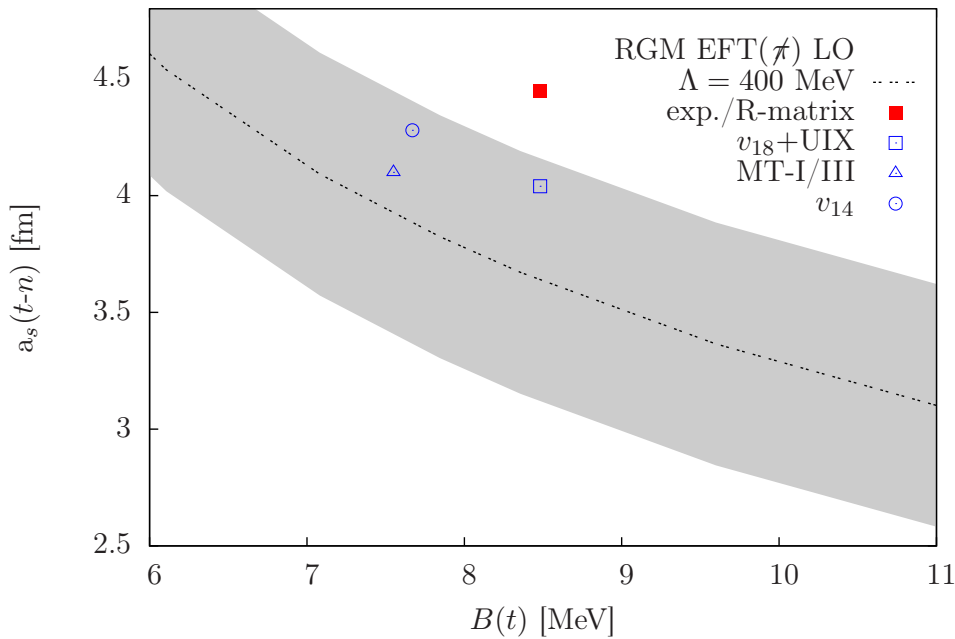}
\caption{\label{fig.tnphillips} The singlet S-wave neutron triton scattering length as a function of the triton binding energy.
The correlations for \eftnopi~with regulators from $400~$MeV to $1.6~$GeV are compared to R-matrix data~\cite{rmat-tn} (red solid lines), and
to the MT-I/III~\cite{mti-iii-bt}, $v_{14}$~\cite{v14-bt}, and the $v_{18}+$UIX~\cite{uix} force models with $\ants$ from~\cite{laz-atn}.}
\end{figure}

In fig.~\ref{fig.tnphillips}, the above mentioned correlation between $B(t)$ and $\ants$ is shown to resolve the discrepancies in
the theoretical predictions.
For each cutoff, the TNI parameter is varied generating the correlation lines (solid black for $\Lambda=400~$MeV).
The lines map out a band (gray) whose width at fixed $B(t)$ represents the LO uncertainty of the calculation. Again, $\Lambda=400~$MeV is chosen to generate
the central line.
The model predictions (empty symbols) are consistent with this analysis as they are included in the band.
The correlation explains the difference in model predictions for the four-body scattering length as being largely due to their difference in $B(t)$.
In the sense that a four-body observable is correlated to a three-body observable, the finding is analogous to the Tjon correlation~\cite{tjon}
between the two binding energies $B(t)$ and $B(^4\text{He})$. By correlating a scattering to a bound-state observable, the graph shown in fig.~\ref{fig.tnphillips}
is the analog of the Phillips line (empirical:~\cite{phillips-line}, EFT explanation:~\cite{tni-gang-3,platter-three-nucl-nlo}).
\par
It is worth mentioning as an outlook that, like the Tjon line, the above correlation is support for a conjecture formulated in~\cite{braham-conjecture}
that the parameters of quantum chromo dynamics place it close to a critical renormalization-group trajectory. A small variation in those parameters, namely the up and down-quark
masses, would then yield divergent two-nucleon scattering lengths and a universal Efimov spectrum in the three-nucleon system. In this socalled resonant limit, the analogous bosonic
system was shown both theoretically~\cite{platter-hammer-4boson,stecher-4boson} and experimentally~\cite{4boson-exp-1,4boson-exp-2} to have two bound four-body states associated
with each Efimov trimer.
The bound state and a shallow resonance in the Helium-4 $0^+$ channel hint towards an analogous universal four-fermion system emerging from three resonantly interacting nucleons.
Whether or not a bound Hydrogen-4 and/or Lithium-4 nucleus will emerge in the limit of diverging two-nucleon scattering lengths would therefore further the understanding
of atomic and nuclear systems within the same theoretical framework.
\par
%--
\begin{figure}
  \includegraphics[width=.97\columnwidth]{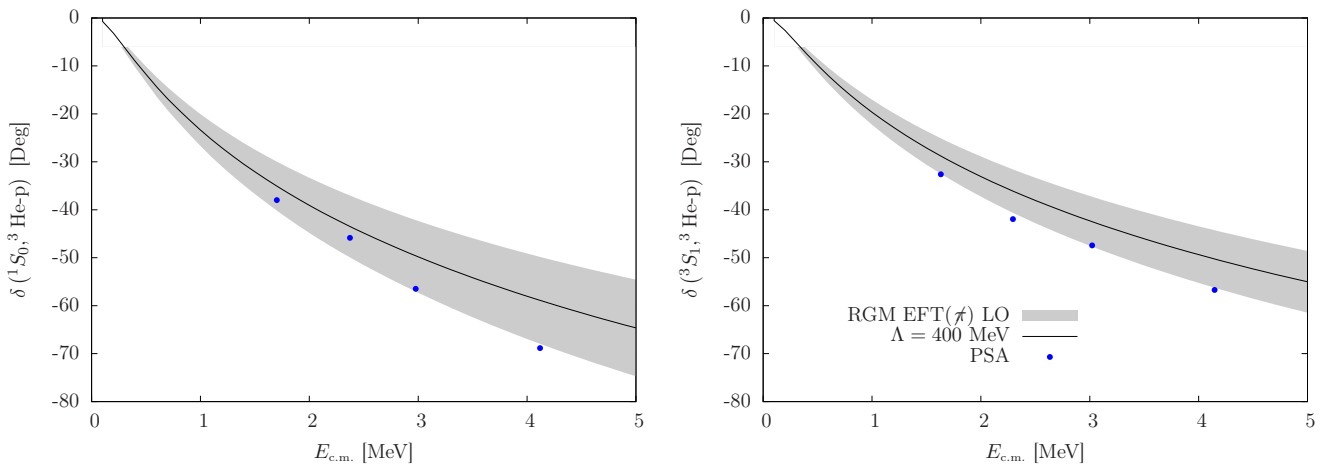}
\caption{\label{fig.hep-phases} Singlet (left) and triplet (right) S-wave phase shifts for \phe~scattering. The LO \eftnopi~error band resembles a
cutoff variation between $400~$MeV (solid line, central phase) and $1.6~$GeV (upper edge of shaded band). Included are the most recent PSA values~\cite{phe-psa}~(points).}
\end{figure}

%--
Finally, the predictions of \eftnopi~ for the singlet and triplet S-wave phase shifts for the \phe~system are compared to the PSA of~\cite{phe-psa}
in fig.~\ref{fig.hep-phases}.
The LO results of this work constrain the phases to the plotted band which was mapped out by the cutoff- and input-data variation.
The upper edge of the band corresponds to $\Lambda=1.6~$GeV while the band center is defined with $\Lambda=400~$MeV.
The width of the band increases with energy,
consistent with the expectation of a larger uncertainty due to a slower convergence of the EFT at higher energies.
Note that if the lower edge is used as the central phase shift, the PSA values lie within the uncertainty range.
No significant shift in the phase shifts over the considered energy range was observed when the proton-proton
scattering length was fitted by considering charge-symmetry-breaking interactions.
The spread is induced predominantly by the varying $B(^3\text{He})$ as only $B(t)$ is held fixed.
%--
%----------------------------------------------------------------------------------------------------------------------------------------------------
\section{Summary}
Model-independent constraints for the singlet and triplet S-wave proton-Helium-3 and neutron-triton scattering lengths were calculated.
The strong dependence of the predictions on the binding energy of the respective three-nucleon target
was quantified by a calculation of a leading-order correlation between the triton binding energy and $\ant$. On the other hand, the dependence of the
four-nucleon scattering lengths on their two-nucleon siblings $\ann$~and $\app$~was found relatively weak.
The effective-field-theory approach ensured consistency of the predictions, first, with relevant two- and three-body data, and second,
with the most recent phase-shift analysis in the \phe~system. A puzzling underprediction of parameters remains in the \nt~case, which is also found when sophisticated models
are used for the prediction of \nt~scattering lengths.
%--
\section{acknowledgments}
The author acknowledges the numerous discussions with Daniel Phillips. This research was supported by the US Department of Energy under grant
DE-FG02-93ER40756.
%--
\bibliographystyle{unsrt}
\bibliography{gen-bib}

\begin{thebibliography}{10}

\bibitem{gfmc}
S.C. Pieper and R.B. Wiringa.
\newblock {\em Ann. Rev. Nucl. Part. Sci.}, 51:53--90, 2001.

\bibitem{lit}
W.~Leidemann.
\newblock {\em Int. J. Mod. Phys. E}, 18:1339--1358, 2009.

\bibitem{hmh-rrgm}
H.M. Hofmann.
\newblock In L.S. Ferreira, A.C. Fonseca, and L.~Streit, editors, {\em
  Proceedings of Models and Methods in Few-Body Physics, Lisboa, Portugal},
  page 243, 1986.

\bibitem{fy}
R.~Lazauskas and J.~Carbonell.
\newblock {\em Few Body Syst.}, 34:105--111, 2004.

\bibitem{uix}
B.S. Pudliner, V.R. Pandharipande, J.~Carlson, S.C. Pieper, and R.B. Wiringa.
\newblock {\em Phys. Rev. C}, 56:1720--1750, 1997.

\bibitem{bonn}
R.~Machleidt, K.~Holinde, and C.~Elster.
\newblock {\em Phys. Rept.}, 149:1--89, 1987.

\bibitem{eft-mam-rev}
S.R. Beane, P.F. Bedaque, W.C. Haxton, D.R. Phillips, and M.J. Savage.
\newblock {From hadrons to nuclei: Crossing the border}.
\newblock In M.~Shifman, editor, {\em At the frontier of particle physics},
  volume~1, pages 133--269, 2000.
\newblock arXiv:nucl-th/0008064.

\bibitem{rev-pionless-platter}
L.~Platter.
\newblock {\em Few Body Syst.}, 46:139--171, 2009.

\bibitem{platter-4bdy-lo}
L.~Platter, H.-W. Hammer, and U.-G. Mei{\ss}ner.
\newblock {\em Phys. Lett. B}, 607:254--258, 2005.

\bibitem{bauchpinsel}
J.~Kirscher, H.W. Grie{\ss}hammer, D.~Shukla, and H.M. Hofmann.
\newblock {\em Eur. Phys. J. A}, 44:239--256, 2010.

\bibitem{phe-psa}
T.V. Daniels, C.W. Arnold, J.M. Cesaratto, T.B. Clegg, A.H. Couture, H.J.
  Karwowski, and T.~Katabuchi.
\newblock {\em Phys. Rev. C}, 82:034002, 2010.

\bibitem{phe-knutson}
E.A. George and L.D. Knutson.
\newblock {\em Phys. Rev. C}, 67:027001, 2003.

\bibitem{phe-data8}
M.T. Alley and L.D. Knutson.
\newblock {\em Phys. Rev. C}, 48:1901, 1993.

\bibitem{n-d-wit-nog}
H.~Witala, A.~Nogga, H.~Kamada, W.~Gl\"ockle, J.~Golak, and R.~Skibinski.
\newblock {\em Phys. Rev. C}, 68:034002, 2003.

\bibitem{phillips-line}
A.C. Phillips.
\newblock {\em Nucl. Phys. A}, 107:209, 1968.

\bibitem{platter-tjon}
L.~Platter, H.-W. Hammer, and U.-G. Mei{\ss}ner.
\newblock {\em Phys. Lett. B}, 607:254--258, 2005.

\bibitem{tjon}
J.A. Tjon.
\newblock {\em Phys. Lett. B}, 56:217, 1975.

\bibitem{walz}
M.~Walzl, U.G. Mei{\ss}ner, and E.~Epelbaum.
\newblock {\em Nucl. Phys. A}, 693:663, 2001.

\bibitem{hmh-4he}
H.M. Hofmann and G.M. Hale.
\newblock {\em Phys. Rev. C}, 77:044002, 2008.

\bibitem{gen-ere}
L.P. Kok, J.W. de~Maag, H.H. Brouwer, and H.~van Haeringen.
\newblock {\em Phys. Rev. C}, 26:6, 1982.

\bibitem{GO99}
D.E.~Gonz{\'a}lez et~al.
\newblock {\em Phys. Rev. Lett.}, 83:3788, 1999.

\bibitem{GO06}
D.E.~Gonz{\'a}lez et~al.
\newblock {\em Phys. Rev. C}, 73:034001, 2006.

\bibitem{HU001}
V.~Huhn et~al.
\newblock {\em Phys. Rev. Lett.}, 85:1190, 2000.

\bibitem{HU002}
V.~Huhn et~al.
\newblock {\em Phys. Rev. C}, 63:014003, 2000.

\bibitem{laz-atn}
R.~Lazauskas and J.~Carbonell.
\newblock {\em Few Body Syst.}, 34:105--111, 2004.

\bibitem{nt-phe-kiev}
M.~Viviani, S.~Rosati, and A.~Kievsky.
\newblock {\em Phys. Rev. Lett.}, 81:1580, 1998.

\bibitem{nt-carbonell}
J.~Carbonell.
\newblock {\em Nucl. Phys. A}, 684:281c, 2001.

\bibitem{phe-data10}
M.~Viviani, A.~Kievsky, S.~Rosati, E.A. George, and L.D. Knutson.
\newblock {\em Phys. Rev. Lett.}, 86:3739, 2001.

\bibitem{phe-data6}
H.~Berg, W.~Arnold, E.~Huttel, H.H. Krause, J.~Ulbricht, and G.~Clausnitzer.
\newblock {\em Nucl. Phys. A}, 334:21, 1980.

\bibitem{phe-data-fisher}
B.M.~Fisher et~al.
\newblock {\em Phys. Rev. C}, 74:034001, 2006.

\bibitem{rmat-tn}
G.M.~Hale et~al.
\newblock {\em Phys. Rev. C}, 42:438, 1990.

\bibitem{nt-exp}
H.~Rauch, D.~Tuppinger, H.~Woelwitsch, and T.~Wroblewski.
\newblock {\em Phys. Lett. B}, 165:39, 1985.

\bibitem{tni-gang-3}
P.F. Bedaque, G.~Rupak, H.W. Grie{\ss}hammer, and H.-W. Hammer.
\newblock {\em Nucl. Phys. A}, 714:589--610, 2003.

\bibitem{platter-three-nucl-nlo}
L.~Platter.
\newblock {\em Phys. Rev. C}, 74:037001, 2006.

\bibitem{braham-conjecture}
E.~Braaten and H.-W. Hammer.
\newblock {\em Phys. Rev. Lett.}, 91:102002, 2003.

\bibitem{platter-hammer-4boson}
H.-W. Hammer and L.~Platter.
\newblock {\em Eur. Phys. J. A}, 32:113, 2007.

\bibitem{stecher-4boson}
J.~von Stecher, J.P. D{'}Incao, and C.H. Greene.
\newblock {\em Nature Phys.}, 5:417, 2009.

\bibitem{4boson-exp-1}
F.~Ferlaino et~al.
\newblock {\em Phys. Rev. Lett.}, 102:2009, 140401.

\bibitem{4boson-exp-2}
D.~Dries S.E.~Pollack and R.G. Hulet.
\newblock {\em Science}, 326:2009, 1683.

\bibitem{mti-iii-bt}
M.R. Hadizadeh, L.~Tomio, and S.~Bayegan.
\newblock {\em arXiv e-prints}, 2011.

\bibitem{v14-bt}
A.~Picklesimer, R.A. Rice, and R.~Brandenburg.
\newblock {\em Phys. Rev. C}, 45:5, 1992.

\end{thebibliography}
\end{document}